\documentclass[fleqn,12pt,twoside]{article}
\usepackage{espcrc1}

\usepackage{graphicx}
\usepackage[figuresright]{rotating}

\newcommand{\AmS}{{\protect\the\textfont2
  A\kern-.1667em\lower.5ex\hbox{M}\kern-.125emS}}

\title{Abundance Anomalies in NGC6752 - Do AGB Stars Have a Role?}

\author{S. W. Campbell\address[CSPA]{Centre for Stellar and Planetary Astrophysics, Monash University, \\
        Melbourne, Australia}
        Y. Fenner\address[SWIN]{Centre for Astrophysics and Supercomputing, Swinburne University of Technology, Melbourne, Australia}
        A. I. Karakas\address{ICA, Department of Astronomy and Physics, Saint Mary's University, Halifax, Canada}
        J. C. Lattanzio\addressmark[CSPA]
        and
        B. K. Gibson\addressmark[SWIN]}
       
\begin{document}

\maketitle

\begin{abstract}
        
We are in the process of testing a popular theory that the observed abundance anomalies in the Globular Cluster NGC 6752 are due to `internal pollution' from intermediate mass asymptotic giant branch stars. To this end we are using a chemical evolution model incorporating custom-made stellar evolution yields calculated using a detailed stellar evolution code. By tracing the chemical evolution of the intracluster gas, which is polluted by two generations of stars, we are able to test the internal pollution scenario in which the Na- and Al-enhanced ejecta from intermediate mass stars is either accreted onto the surfaces of other stars, or goes toward forming new stars.  In this paper we focus mainly on the nucleosynthetic yields of the AGB stars and discuss whether these stars are the source of the observed Na-O anticorrelation. Comparing our preliminary results with observational data suggests that the qualitative theory is not supported by this quantitative study. 

This study has recently been completed and published in \cite{FC04}. Details of the stellar models will be in a forthcoming paper \cite{CA04}.

\end{abstract}

\section{Introduction}

As with most globular clusters (GCs), NGC 6752 has a very uniform distribution of heavy elements (see eg. \cite{BH84}). This implies that there was thorough mixing of stellar ejecta with the ambient gas during the early history of the cluster.
 Conversely, most GCs show large abundance variations in the light elements (eg. the C-N, Mg-Al and O-Na anticorrelations, eg. \cite{GR01}). This indicates an inhomogeneous source of pollutants must have been produced after the 'well-mixed' era.
As heavy elements are primarily produced by massive stars, the evidence suggests that there was a generation of massive stars which polluted the pristine protocluster gas, followed by a later release of lighter elements, presumably from lower mass stars.
In our model we assume a two stage star formation/pollution history, similar to that used in the dynamical evolution study by Parmentier et al. \cite{PJ99}.

\section{The Pollution Scenario}

\subsection{Stage 1 Pollution}

The mass distribution used to model the first stars that pollute the primordial gas was based on the work of Nakamura \& Umemura \cite{NU01}. They predict a bimodal initial mass function (IMF) for a Z=0 population. In addition to being bimodal, it is also 'top-heavy', favouring massive star formation.

Stellar yields were taken as input for the chemical evolution (CE) model. The yields of \cite{UN02} were used for the 150-270 M$_{\odot}$ range, ~\cite{CL02} for the 13-80 M$_{\odot}$ range, and Karakas (2003, priv. comm.) for the intermediate mass stars \cite{KL03} (nb. these were Z=0.0001 models but played a negligible role in polluting the early cluster due to the top-heavy IMF). 
Star formation occurred in a single burst, with newly synthesised elements returned on timescales prescribed by mass dependent lifetimes \cite{GM01}. In this way the cluster gas was brought up to [Fe/H] = $-1.4$ , the current value (e.g. \cite{GR01}), from which the next generation of stars formed.

\subsection{Stage 2 Pollution}

Stage 2 in the model sees a population of stars forming from a mix of the ejecta of the Z=0 stars and Big Bang material.
A standard IMF \cite{KT93} was adopted for stage 2. However, we assumed that the GC only retained the ejecta from stars with  mass $< 7 M_{\odot}$ --- the winds and ejecta from SNe are assumed to have escaped the system due to their high velocities. Thus, only the yields from intermediate mass stars impact upon the chemical evolution from then on. 
Yields from a specifically calculated grid of models were used as self-consistent feedback in the CE model. They are described below.

\section{Second Generation AGB Stars}

\subsection{The Fiducial Set of Models}

A custom-made set of low- and intermediate-mass stars were computed for the second epoch of star formation in the cluster, using the Mount Stromlo Stellar Structure Code (eg. see \cite{FL96}) and the Monash University Stellar Nucleosynthesis Code. This generation of stars has a unique chemical composition given by the first stage of pollution - they lack nitrogen, are $\alpha$-enhanced, and are low metallicity. The models \cite{CA04} were computed using the exact (non-scaled-solar) composition that resulted from the evolution of the chemical pollution model described above. This required removing all scaled-solar assumptions from the codes and computing new opacity tables specifically for these stars (OPAL tables were used, eg. \cite{IR96}).
For this `fiducial' set of models, Reimers' mass-loss law \cite{RE75} was used on the red giant branch (RGB) and Vassiliadis \& Wood's law \cite{VW93} on the asymptotic giant branch (AGB). Nuclear reaction rates were mainly taken from the REACLIB Data Tables \cite{TA91}.

\subsection{Yields and The Effects of Switching Reaction Rate Compilations \& Mass Loss Prescriptions}

\begin{figure}[!htb]
\begin{minipage}[!t]{83mm}
\includegraphics[width=86mm,angle=0]{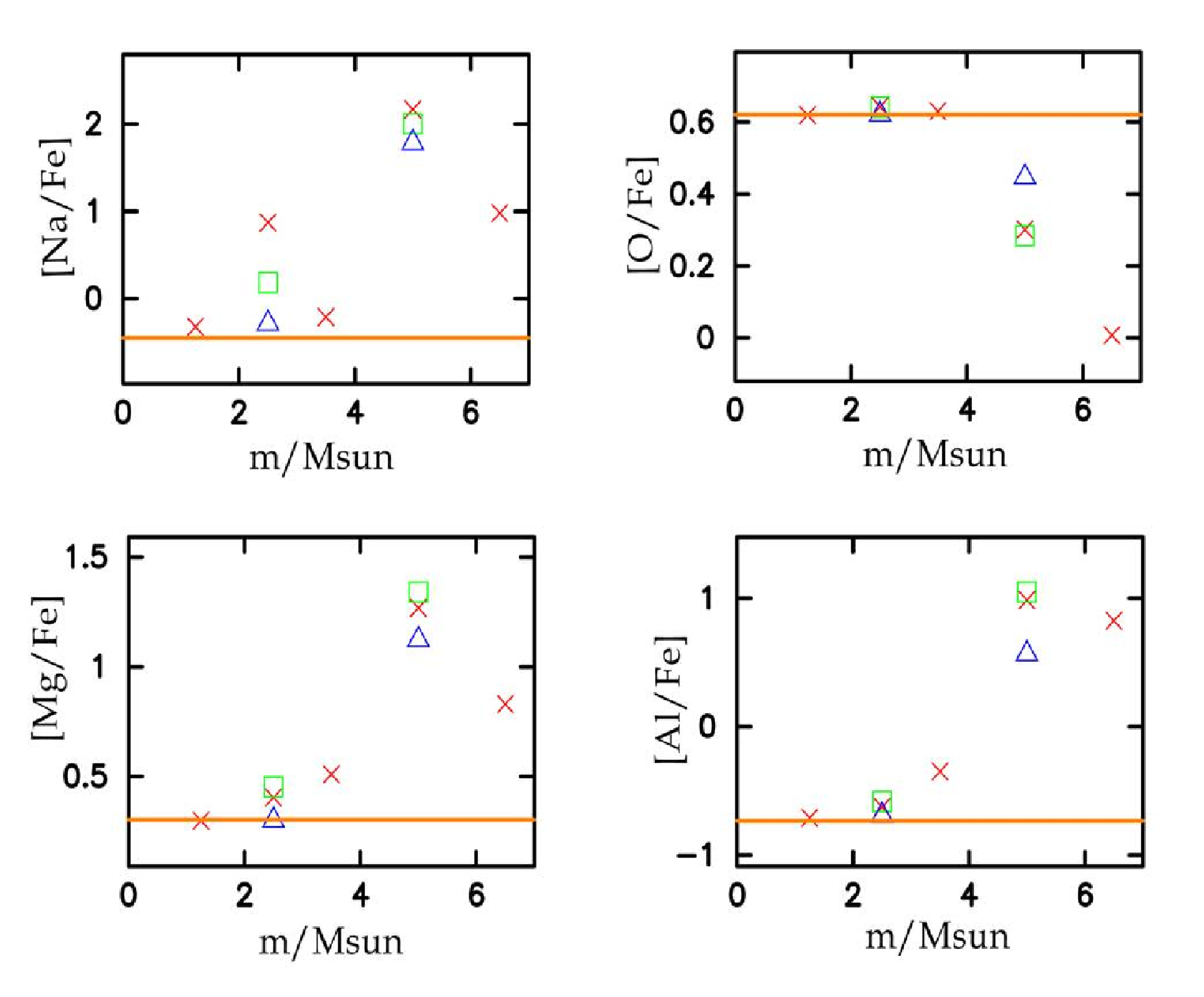}
\caption{Comparison of stellar model yields for selected elements. The fiducial set (crosses) uses Vassiliadis \& Wood's (VW) mass-loss rate on the AGB and the fiducial nuclear reaction rates. The squares represent yields for the case where the NACRE rates were used in combination with the VW mass-loss (for masses 2.5 \& 5 M$_{\odot}$ only). Triangles represent yields from models with our fiducial rates but use Reimers' mass-loss formula on the AGB (2.5 \& 5 M$_{\odot}$). The solid line is the initial composition of the stars.}
\label{yieldsfig}
\end{minipage}
\hspace{\fill}
\begin{minipage}[!t]{74mm}
\includegraphics[width=73mm,angle=0]{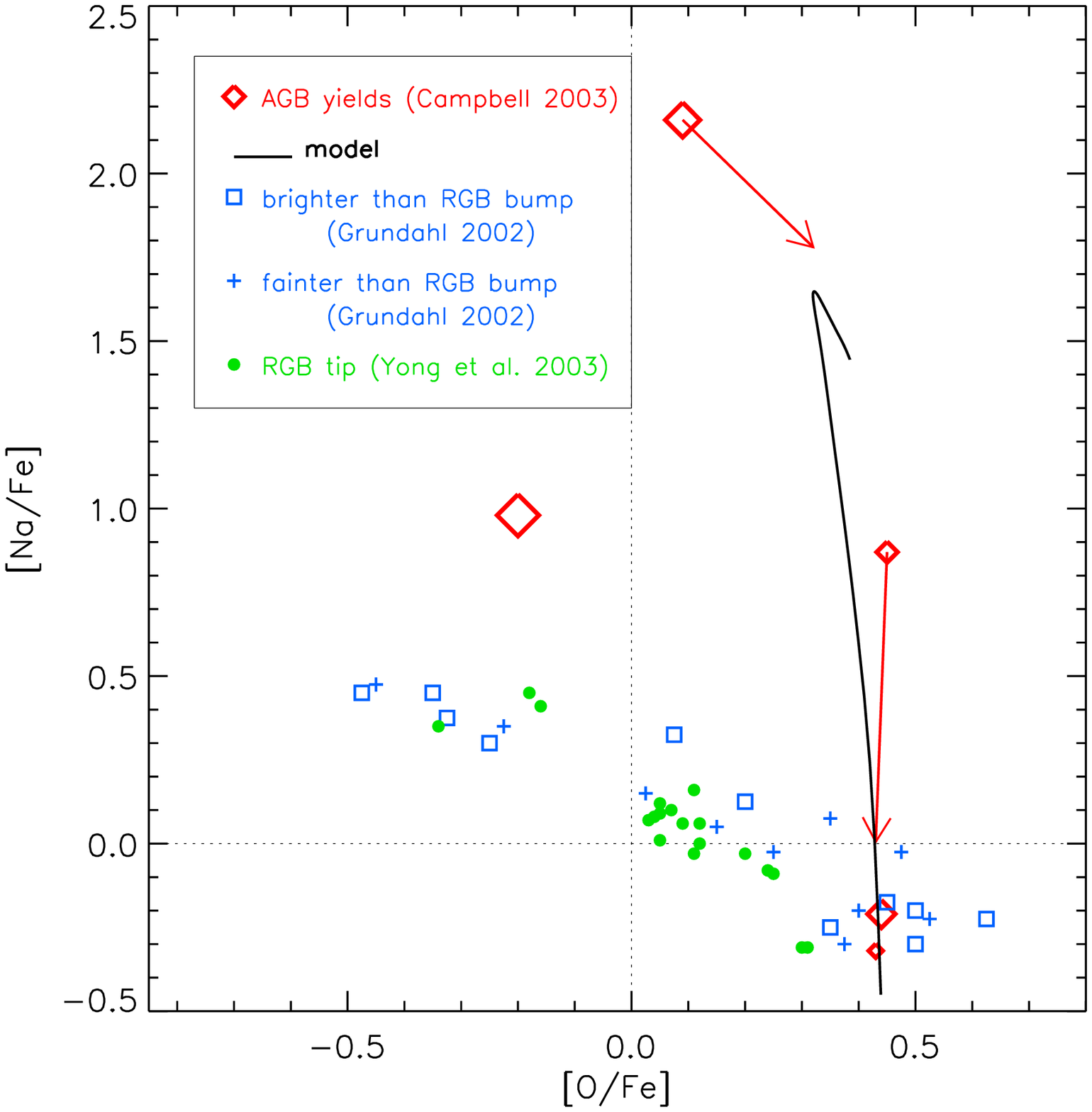}
\caption{The O-Na anticorrelation in NGC 6752. The solid line is the predicted trend given by tracking the chemical evolution of the intracluster medium. Dots, crosses and squares are the observational data, while the diamonds are the AGB stellar model results (diamonds increase in size with stellar mass; 1.25, 2.5, 3.5, 5.0, 6.5 M$_{\odot}$). The arrows indicate the change due to the use of Reimers' mass-loss law on the AGB (instead of VW). (Observational data references: ~\cite{GB02,YG03}).}
\label{NaOfig}
\end{minipage}
\end{figure}

To estimate the sensitivity of the GC chemical evolution model to the various prescriptions used in the stellar model calculations, some experiments were run in which the following were altered: 1) the reaction rate compilation used for the NeNa chain,  MgAl chain and $^{22}$Ne + $\alpha$ reactions  and 2) the mass-loss formula used for the AGB evolution. 
Reaction rates and mass-loss are two of the key uncertainties in the stellar models. NACRE rates \cite{AA99} were used as the alternative compilation, and the mass loss laws were altered for AGB evolution (see Fig.~\ref{yieldsfig} caption for details). Fig.~\ref{yieldsfig} shows the results of both tests as compared to the yields from the fiducial set of models.
It was found that altering the mass-loss formula on the AGB had a greater impact on the stellar yields than altering the reaction rates. The effects of these changes are further diluted when used in the CE model.

\section{Results}

Fig.~\ref{NaOfig}  compares the GCCE model results with observational data for the O-Na anticorrelation. The predicted variations in Na and O do not match the observations.
The spread in Na is easily achieved (actually too much is produced), but the oxygen is not depleted enough. Although O is significantly depleted in the 5.0 and 6.5 M$_{\odot}$ stellar models, through Hot Bottom Burning (HBB) on the AGB, these stars are not very numerous in a standard IMF.

\section{Summary}

Using detailed nucleosynthetic yields we have computed the chemical evolution of the intracluster medium in the globular cluster NGC 6752. Abundance spreads in Na and O were found, however too much Na was produced, while the O is not sufficiently depleted to account for the observations --- the Na-O anticorrelation was not matched. Altering the IMF to favour more massive AGB stars may allow the matching of this anticorrelation but may adversely effect the other elements. Thus, although the second generation intermediate-mass AGB stars do show the hot hydrogen burning (via HBB) that is required to explain the observations, this quantitative study suggests that HBB in AGB stars may {\em not} actually be the site. Changing the mass-loss prescriptions and reaction rate compilations in the stellar models does not alter this conclusion.

This work has recently been completed and published in \cite{FC04}, in which the Mg-Al correlation as well as predictions for C, N and He are discussed.


\begin{thebibliography}{9}

\bibitem{FC04} Fenner, Y., Campbell, S.W., Karakas, A.I., Lattanzio, J.C, Gibson, B.K., 2004, MNRAS, 353, 789
\bibitem{CA04} Campbell, S. W., et al. 2004, in prep.
\bibitem{BH84} Bell, R. A., Hesser, J. E., \& Cannon, R. D., 1984, ApJ, 283, 615B
\bibitem{GR01} Gratton, R. G. et al., 2001, A\&A 369, 87
\bibitem{PJ99} Parmentier, G., Jehin, E., Magain, P., et al., 1999, A\&A, 352, 138
\bibitem{NU01} Nakamura, F. \& Umemura, M., 2001, ApJ 548, 19
\bibitem{UN02} Umeda, H. \& Nomoto, K., 2002, ApJ 565, 385
\bibitem{CL02} Chieffi, A. \& Limongi, M., 2002, ApJ 577, 281
\bibitem{KL03} Karakas, A. I. \& Lattanzio, J. C. 2003, PASA, 20, 279
\bibitem{GM01} Gusten, R. \& Mezger, P. G., 1982, Vistas Astron. 26, 159
\bibitem{KT93} Kroupa, P., Tout, C. A., \& Gilmore, G. 1993, MNRAS, 262, 545
\bibitem{FL96} Frost, C. A. \& Lattanzio, J. C. 1996, ApJ, 473, 383
\bibitem{IR96} Iglesias C.A., Rogers F.J., 1996, ApJ, 464, 943
\bibitem{RE75} Reimers, D. 1975, Memoires of the Societe Royale des Sciences de Liege, 8, 369
\bibitem{VW93} Vassiliadis, E. \& Wood, P. R. 1993, ApJ, 413, 641
\bibitem{TA91} Thielemann, F., Arnould, M. \& Truran, J. W. 1991, in Advances of Nuclear Astrophysics, ed. E. Vangioni-Flam et al. (France: Editions Frontieres), 525
\bibitem{AA99} Angulo C., Arnould M., Rayet M., et al., 1999, Nuclear Physics A 656, 3
\bibitem{GB02} Grundahl, F., Briley, M, Nissen, P. E. \& Feltzing, S. 2002, A\&A 385, L14
\bibitem{YG03} Yong, D., Grundahl, F., Lambert, D. L., et al. 2003, A\&A, 402, 985

\end{thebibliography}
\end{document}